\documentclass[10pt,a4paper]{article}

\RequirePackage[OT1]{fontenc}
\RequirePackage{amsthm,amsmath}
\RequirePackage[authoryear,round]{natbib}
\usepackage[a4paper,hyperindex,
            colorlinks=true,linktocpage=false,
            bookmarks=false,
            breaklinks=true,
            pdftitle={Random-effects meta-analysis of Phase I dose-finding studies},
            pdfauthor={Ursino, M. et al.},
            pdfkeywords={},
            linkcolor=black,citecolor=black,urlcolor=black]{hyperref}

\RequirePackage{xcolor}
\usepackage{multirow}
\usepackage{graphicx}
\usepackage{booktabs}      % for tables
\usepackage{dsfont}        % for `\realline'

\DeclareMathOperator*{\argmin}{arg\,min} 
\providecommand{\binomdistn}{\mathrm{Binomial}}
\providecommand{\betadistn}{\mathrm{Beta}}
\providecommand{\gammadistn}{\mathrm{Gamma}}
\providecommand{\normaldistn}{\mathrm{Normal}}

\providecommand{\halfnormaldistn}{\mathrm{Halfnormal}}
\providecommand{\realline}{\mathds{R}}
\DeclareMathOperator{\logit}{logit}
\DeclareMathOperator{\expect}{E}
\DeclareMathOperator{\var}{Var}
\DeclareMathOperator{\prob}{P}
\providecommand{\sigmam}{\sigma_{\mathrm{m}}}

\title{Random-effects meta-analysis\\ of phase~I dose-finding studies\\ using stochastic process priors}

\author{Moreno Ursino\textsuperscript{1,2,}\thanks{Email: \texttt{moreno.ursino@inserm.fr}},
        Christian R\"{o}ver\textsuperscript{3}, Sarah Zohar\textsuperscript{1}, and Tim Friede\textsuperscript{3}}
\date{\textsuperscript{1} Inserm, Paris, France \\
      \textsuperscript{2} F-CRIN Partners Platform, Paris, France \\
      \textsuperscript{3} University Medical Center G\"{o}ttingen, G\"{o}ttingen, Germany \\[3ex]
      August 1, 2019}

\begin{document}

\maketitle

\begin{abstract}
  Phase~I dose-finding studies aim at identifying the maximal
  tolerated dose (MTD). It is not uncommon that several dose-finding
  studies are conducted, although often with some variation in the
  administration mode or dose panel. For instance, sorafenib (BAY
  43-900) was used as monotherapy in at least 29 phase~I trials
  according to a recent search in \href{https://www.clinicaltrials.gov/}{\texttt{clinicaltrials.gov}}. Since the
  toxicity may not be directly related to the specific indication,
  synthesizing the information from several studies might be
  worthwhile. However, this is rarely done in practice and only a
  fixed-effect meta-analysis framework was proposed to date.  We
  developed a Bayesian random-effects meta-analysis methodology to
  pool several phase~I trials and suggest the MTD\@. A curve free
  hierarchical model on the logistic scale with random effects,
  accounting for between-trial heterogeneity, is used to model the
  probability of toxicity across the investigated doses. An
  Ornstein-Uhlenbeck Gaussian process is adopted for the random
  effects structure. Prior distributions for the curve free model are
  based on a latent Gamma process.  An extensive simulation study
  showed good performance of the proposed method also under model
  deviations. Sharing information between phase~I studies can improve
  the precision of MTD selection, at least when the number of trials
  is reasonably large.
\end{abstract}

\section{Introduction}\label{sec:intro}

  Phase~I dose-finding studies are carried out during early stages of
  the clinical development, and aim at estimating the \emph{maximum
    tolerated dose (MTD)} of a drug or a combination of molecules.
  The MTD is defined with reference to the occurrence of
  treatment-related adverse events, so-called \emph{dose-limiting
    toxicities (DLTs)}. The MTD then is reached once the rate of DLTs
  exceeds an acceptable level.  Phase~I studies are usually done on
  small numbers of healthy volunteers, except in oncology, where, due
  to the potentially high toxicity of drugs, phase~I trials are
  commonly performed on patients \citep{chevret06}.

  In oncology, identifying the correct or reasonable dose or set of
  doses is a crucial objective in the drug development process:
  selecting too high a dose means exposing patients to an unacceptable
  toxicity profile, while selecting a dose of too low toxicity
  increases the likelihood that the treatment provides insufficient
  efficacy \citep{bretz05}. The dose escalation paradigm in phase~I
  (or~I/II) trials thus generally aims to avoid recommending too toxic
  doses of an agent while maintaining an acceptable toxicity. Due to
  limited sample sizes, conventional statistical methods are often
  inaccurate, so that adaptive sequential analyses have been proposed,
  as these can potentially find the MTD sooner and limit the number of
  exposed subjects \citep{letourneau09,NeuenschwanderEtAl2015}.

  When combining data across trials, two sources of potential
  heterogeneity need to be considered. Firstly, these are differences
  in the outcomes of the control groups. In the context of
  dose-escalation studies, there might be differences in the (true)
  toxicity probabilities due to variations in e.g. the study
  populations or in the definition and assessment of
  toxicities. Secondly, the (true) treatment effects, even if defined
  on a relative scale, might vary across trials. In standard
  meta-analysis models, the former is addressed by stratification for
  study. In so-called random-effects meta-analyses, the latter is
  addressed by inclusion of random study-by-treatment interactions. In
  fixed-effect or common-effect meta-analysis, a homogeneous treatment
  effect across trials is assumed. For a recent discussion of the
  various statistical models we refer here to
  \cite{JacksonEtAl2018}. As evidenced by large-scale empirical
  investigations, some level of between-study heterogeneity is not
  unlikely to occur \citep{TurnerEtAl2012}. However, estimation of the
  corresponding variance component and accounting appropriately for
  the uncertainty in estimation in inference of relevant model
  parameters can be challenging, if the number of studies included in
  the meta-analysis is small
  \citep{FriedeRoeverWandelNeuenschwander2017}. In the context of
  meta-analyses of dose-escalation trials, we are still lacking an
  understanding as well as empirical evidence how the various forms of
  between-trial heterogeneity can be appropriately accounted for.
  
  \citet{zohar11} proposed a meta-analysis approach for phase~I
  clinical trials in oncology. Phase~I data were pooled while
  accounting for the sequential nature of such trials in order to
  better estimate the overall MTD\@. However, this method did not deal
  with several important characteristics associated with phase~I
  features. Firstly, data were pooled under several different
  administration schedules, which may imply different toxicity
  profiles. Secondly, between-trial heterogeneity was not taken into
  account, which may lead to inaccurate inference. Thirdly, as the
  pooled analysis was done retrospectively, it would have been
  possible to take into account cycles, dose-modifications and long
  term toxicities in order to better investigate the maximal dose
  regimen, but these complexities were not addressed.

  \citet{thomas14} reported the results of a meta-analysis based on
  dose-response studies conducted by a large pharmaceutical company
  between 1998 and 2009. Data collection targeted efficacy endpoints,
  but safety data were not extracted. The goal of this meta-analysis
  was to identify consistent quantitative patterns in dose-response
  across different compounds and diseases. The meta-analysis excluded
  oncology trials as these have different dosing objectives and
  methods.

  In this manuscript, we develop a novel meta-analysis approach for
  phase~I clinical trials in oncology, which takes into account the
  different features described above to better suit the requirements
  in estimating MTDs. We generalized the binomial-normal hierarchical
  model (BNHM), which is most commonly used in the literature for
  meta-analysis of studies involving a single dose.  In the following
  section, two motivating examples are described. In
  Section~\ref{sec:methods}, the methodology is presented, along with
  prior distributions and different variations of MTD definitions.  In
  Section~\ref{sec:simul} we describe model variations and simulation
  settings that we used to test the developed method and its
  sensitivity to varying circumstances.  Finally, in
  Section~\ref{sec:application}, the new methodology is applied to the
  motivating case studies and some limitations are discussed in
  Section~\ref{sec:discussion}.

\section{Motivating examples}\label{sec:motivation}

  Some might believe that there are more phase~III than phase~I
  studies, and so meta-analyses have largely focused on late-stage
  trials whereas opportunities in pooling phase~I results have rarely
  been investigated.  Furthermore, as phase~I studies usually have
  small sample sizes and are mostly algorithm-based and only lately
  model based designs, methodologists have been less inclined to
  embrace this issue.  The first illustration concerns sorafenib (BAY
  43-9006) which is a kinase inhibitor approved for the treatment of
  advanced renal cell carcinoma, hepatocellular carcinoma, and
  radioactive iodine resistant advanced thyroid carcinoma. A search of
  the \texttt{clinicaltrials.gov} registry of clinical trials at the
  end of June 2019 revealed that there are at least 833~studies using
  sorafenib (at any recruitment stage and type of study) of which
  248~studies were labeled as ``phase~I'' or ``phase~I/II'' and
  99~studies were labeled as ``phase~III'' or ``phase~II/III''. Of 248
  phase~I or phase~I/II studies using sorafenib, 29~studies used it in
  phase~I as monotherapy (median sample size~22, range 2--158).

  Today, the dose recommended by the European Medicines Agency (EMA)
  is 400 milligrams (mg) twice a day. Several phase~I studies on
  sorafenib monotherapy have been performed, and some of their results
  are summarized in Table~\ref{tab:sorafenib}.  Within these 14
  trials, a total of 7~doses were tested, with most of these studies
  targeting solid tumors or leukemia. DLT definitions were comparable,
  and most of sorafenib schedules followed a 28-day cycle or similar.
  
  \begin{table}
    \caption{The results of 14 studies on sorafenib monotherapy. For
      each dose considered in each trial, the numbers of patients
      experiencing dose-limiting toxicities events, and the total numbers of exposed
      patients are given.}
    \label{tab:sorafenib}
    \begin{tabular}{l ccccccc}
      \toprule
      & \multicolumn{7}{c}{Dose (mg)}  \\
      \cmidrule(lr){2-8}
      Study          & 100 &  200 & 300 &  400 &  600 & 800 & 1000 \\
      \midrule
      \citet{clark05}       & 0/3 &  0/3 &     &  1/4 &  1/6 & 3/3 & \\
      \citet{awada05}       & 0/4 &  0/3 & 1/5 & 1/10 & 7/12 & 1/3 & \\
      \citet{moore05}       & 0/3 &  1/6 &     &  0/8 &  3/7 &     & \\
      \citet{strumberg2005} & 1/5 &  1/6 &     & 0/15 & 4/14 & 2/7 & \\
      \citet{minami08}      & 0/3 & 1/12 &     &  0/6 &  1/6 &     & \\
      \citet{miller09}      &     & 8/34 &     & 6/20 &      &     & \\
      \citet{nabors11}      &     &  0/3 &     &  1/6 &  0/3 & 1/5 & 3/3 \\
      \citet{chen14}        &     &  0/3 &     & 1/16 &      &     & \\
      \citet{jia13}         &     &      &     &  3/4 &      &     & \\
      \citet{borthakur11}-1 &     &  0/3 &     & 0/15 &  2/8 &     & \\
      \citet{borthakur11}-2 &     &  0/3 &     &  1/7 &  2/6 &     & \\
      \citet{crump10}-1     & 0/4 &  1/6 & 0/6 &  1/6 &      &     & \\
      \citet{crump10}-2     & 0/3 &  1/6 &     &  0/3 &  2/6 &     & \\
      \citet{furuse08}      &     & 0/12 &     & 1/14 &      &     & \\
    \bottomrule
    \end{tabular}
  \end{table}

  Applying the common-effect approach proposed by \citet{zohar11} (in
  the following referred to as the ZKO approach) to the sorafenib data
  (Table~\ref{tab:sorafenib}) and using $(0.05, 0.1,0.2, 0.3, 0.45,
  0.6, 0.65)$ as skeleton, that is the set of prior toxicity
  probabilities for the doses (chosen in a reasonable shape according
  \citet{OQuigley10} and \citet{zohar11}), the following estimated
  toxicity probabilities are obtained: (0.012, 0.033, 0.093, 0.169,
  0.308, 0.471, 0.53). Following the ZKO~approach, and assuming a
  toxicity threshold of 0.33, a dose of 600~mg is estimated as MTD,
  while for a threshold of 0.2, the MTD is at 400~mg.
 
  The second example concerns a combination therapy of irinotecan and
  S-1 (S-1 refers to a combination of three pharmacological compounds,
  namely tegafur, gimeracil, and oteracil potassium).  Irinotecan is a
  topoisomerase~1 inhibitor. It has proven effective in combination
  with 5-fluorouracil (5-FU) but was associated with many adverse
  events. This is why the association with S-1 instead of 5-FU was
  evaluated. In this case 11~studies were used
  (Table~\ref{tab:irinotecanS-1}) in which 10~doses were evaluated
  across all trials.
 
  Applying the ZKO method on the S-1 data
  (Table~\ref{tab:irinotecanS-1}) and using (0.005, 0.05, 0.1, 0.2,
  0.3, 0.4, 0.5, 0.6, 0.65, 0.70) as skeleton we obtain the following
  estimated toxicity probabilities: (0.002, 0.026, 0.061, 0.141,
  0.231,$ $0.328, 0.43, 0.537, 0.592, 0.648). Assuming a toxicity
  threshold of 0.33, dose 90 mg/m$^2$ is estimated as MTD, while for a
  threshold of 0.2, the MTD is at 80 mg/m$^2$.

  \begin{table}
    \caption{The results of 10 studies on combination therapy of
      irinotecan and S-1 (tegafur/gimeracil/oteracil). For each dose
      considered in each trial, the numbers of patients experiencing
      dose-limiting toxicities events, and the total numbers of
      exposed patients are given.}
    \label{tab:irinotecanS-1}
    \begin{tabular}{lcccccccccc}
      \toprule
      & \multicolumn{10}{c}{Dose (mg/m$^2$)}  \\
      \cmidrule(lr){2-11}
      Study & 40 & 50 & 60 & 70 & 80 & 90 & 100 & 120 & 125 & 150 \\
      \midrule
      \citet{ogata09} & 0/3 & 0/3 & 3/4 &  &  &  &  &  &  & \\
      \citet{inokuchi06} &   &   &   & 0/3 & 10/42 & 0/3  & 2/3  &  &  & \\
      \citet{goya12} &   &   &   & 0/3 & 0/3 & 3/5  &  &  &  & \\
      \citet{takiuchi05} & 1/6  &   & 0/3   &  & 0/4 &   & 3/6  &  &  & \\
      \citet{ishimoto09} &   & 0/3  & 0/3   & 0/3 & 2/4 &   &   &  &  & \\
      \citet{kusaba10} &   &    &     &   & 0/6 &   & 2/3  &  &  & \\
      \citet{nakafusa08} &   &    & 7/39    &   & 2/3 &   &   &  &  & \\
      \citet{shiozawa09} &   &    &   &   & 1/6 &   & 2/6  & 2/6 &  & 2/3 \\
      \citet{yoda11} &   &    & 0/3    &   & 3/6 &   &   &  &  & \\
      \citet{komatsu10} &   &    &   &   &  &   &1/9  &  & 1/9  & 0/3 \\
      \bottomrule
    \end{tabular}
  \end{table}

  In the two examples given above, not all trials shared the same
  doses, dose ranges and sample size. The ZKO method was applied to
  estimate the overall MTD\@.  However, this is a simplistic way of
  pooling several adaptive sequential phase~I data sets and it can be
  seen as a fixed-effect meta-analysis method. In the next section
  will be detailed our proposition taking into account these
  specificities as well as inter and intra trial heterogeneity by
  developing a non-parametric random-effects approach.

\section{Methods}\label{sec:methods}
\subsection{The dose-response model}

  In case of studies concerned with only a single dose, the
  binomial-normal hierarchical model (BNHM), or an approximation, is
  most commonly used in the literature
  \citep{JacksonEtAl2018,GuenhanRoeverFriede2018}. When moving to
  several doses in the same study, we propose an extension of the BNHM
  that is adapted to the dose-finding context, and that is able to
  also account for the ordering and spacing among doses.

  Let $k \in \{1, \ldots, K\}$ be the study index, and $i \in \{1,
  \ldots, I\}$ be the dose level index, where all doses~$d_i$ used in
  all trials are indexed in increasing order. Especially with data
  combined from several studies, the dose steps, that is the
  ``spacing'' between neighbouring doses~$d_i$, may be rather
  different (see e.g.\ the irinotecan example in
  Table~\ref{tab:irinotecanS-1}) and needs to be accounted for in the
  model. We define $\delta_{i,j}$ as the metric, specifying the
  spatial proximity or distance between doses.  This may simply be
  defined as the plain difference
  ($\delta_{i,j}=d_i\!-\!d_j$). However, in many cases it may make
  sense to rather consider \emph{relative} differences between dose
  levels on the logarithmic scale
  ($\delta_{i,j}=\log(d_i)\!-\!\log(d_j)=\log\bigl(\frac{d_i}{d_j}\bigr)$).
  Another option may be to assume unit increments for neighbouring
  doses.

  The number of patients in study~$k$ allocated to dose~$i$ is given
  by $n_{ik}$, while $X_{ik}$ is the number of patients experiencing a
  DLT\@.  We then propose the following model:
  \begin{eqnarray}
    X_{ik}           & \sim & \binomdistn \left( n_{ik}, p_{ik} \right) \label{eqn:eventcount} \\
    \logit(p_{ik})   & =    & \sum_{j \leq i} \mu_{j} + b_{ik} \label{eqn:eq_mean} 
  \end{eqnarray}
  where $p_{ik}$~is the probability of toxicity of dose~$i$ in the
  $k$th study.  The probabilities~$p_{ik}$ here are modelled on the
  logit-scale, with $\logit(x)=\log\bigl(\frac{x}{1-x}\bigr)$.

  The \emph{fixed effects} $\mu_{1} \in \realline$ and $\mu_{i} \in
  \realline^+$ (for $i>1$) are common across all studies; the
  summation in \eqref{eqn:eq_mean} ensures non-decreasing overall mean
  probabilities of toxicity with increasing dose.  The \emph{random
    effects} accounting for between-study heterogeneity are
  represented by the (study-specific) vectors~$\mathbf{b}_{k}\sim
  N(\mathbf{0}, \Sigma)$, where $\mathbf{0}$ represents the zero
  vector of dimension $I$ and $\Sigma = \left\lbrace \sigma_{i,j}
  \right\rbrace_{i,j=1, \ldots, I}$ the variance-covariance matrix. In
  order to meaningfully generalize from the BNHM for a single dose to
  a joint model for multiple doses, we specify the fixed and random
  effects accounting for the corresponding dose levels ($d_i$) and
  their ordering and proximity.

\subsection{Gaussian process for the random effects}

  For the random effects, we specify a model that accounts for the
  position of dose~$d_i$ on the dose continuum.  We do not impose
  monotonicity here and we rely on a relatively simple class of
  Gaussian processes.  Two interesting special cases are encompassed
  by the model, namely \emph{independent} and \emph{identical}
  residuals at all doses. In between these extremes, we utilize a
  stationary \emph{Ornstein-Uhlenbeck process (OUP)} with covariance
  \begin{equation}\label{eqn:OupCovariance}
    \textstyle
    \sigma_{i,j}^2 \;=\; \sigmam^2\exp\Bigl(-\frac{|\delta_{i,j}|}{\ell}\Bigr)
  \end{equation}
  where $\sigmam^2$ is the marginal variance, and $\ell>0$ is a
  smoothness parameter determining how quickly the autocorrelation
  decays and residuals become less dependent, depending on the spatial
  separation of doses.  On small scales (relative to~$\ell$), the OUP
  behaves like a Wiener process (or Brownian motion); this nicely
  corresponds with the notion that \emph{if} we knew the residual at a
  certain dose, we knew less about the neighbouring residual the
  further we moved away from that dose, where increments behaved
  (approximately) additively, as for the fixed effects model
  introduced below. For the limiting cases of $\ell \rightarrow 0$ and
  $\ell \rightarrow \infty$ it yields independent or identical
  residuals across doses, respectively
  \citep{uhlenbeck1930,doob42}. Prior distributions for the random
  effect's marginal variance~$\sigmam^2$ and the OUP's spatial
  scale~$\ell$ need to be specified.

\subsection{Gamma process for fixed effects prior distributions}

  The definition of the common effect via a sum of unknown increments
  in \eqref{eqn:eq_mean} places the model into the class of stochastic
  processes, which are commonly used as nonparametric models for
  unkown functions \citep[Ch.~21]{BDA3rd}. Therefore, the prior
  distributions on the unknown increments may be inspired by a
  stochastic process. A natural and convenient class of models is
  defined via \emph{infinitely divisible} probability distributions
  \citep{steutel79}; that means that we stay within the same
  distribution class for the increments (i.e., if we sum two
  increments, the sum's distribution again is within the same
  distribution class), which results in an overall consistent
  model. Since in the present case we are considering strictly
  positive increments for increasing doses, the Gamma process is an
  obvious choice here \citep{lawless04}.

  The Gamma distribution is defined through two parameters, namely the
  \emph{shape}~$k > 0$ and the \emph{scale}~$\theta > 0$; its
  expectation then is $k\theta$ and the variance is
  $k\theta^2$. Choosing the first dose ($d_1$) as the \emph{reference
    dose}, we can specify the prior distributions as a Gamma process
  with
  \begin{eqnarray}
    \mu_1 & \sim & \normaldistn (\mu^\ast, \sigma^\ast), \\
    \mu_i & \sim & \gammadistn (k= \delta^*_{i, i-1}\kappa, \theta) 
                   \quad \mbox{for } i>1, \label{eqn:gamma_prior}
  \end{eqnarray}
  where $\delta^*_{i, i-1}$ is the dose increment from dose~$d_{i-1}$
  to~$d_{i}$. To note, $\delta^*$ can be equal to $\delta$ (used in
  the specification of the random effects), or it can use another
  underlying metric.  The parameter~$\mu_1$ serves as an ``intercept''
  term, and hyperparameters $\mu^\ast$ and $\sigma^\ast$ then need to
  be specified with reference to the expected toxicity at the
  reference dose.  The Gamma process hyperparameters $\kappa$
  and~$\theta$ also need to be pre-specified.  For a sensible choice,
  it is convenient to consider their effect on the conditional
  distribution for a unit increment:
  \begin{eqnarray}
    \expect[\mu_i\,|\,\delta_{i,i-1}\!=\!1] & = & \kappa \theta \\
    \var(\mu_i\,|\,\delta_{i,i-1}\!=\!1)    & = & \kappa \theta^2
  \end{eqnarray}
  which suggests a re-parametrisation in terms of
  \begin{eqnarray}
    \mbox{slope } a                     & = & \kappa \theta \quad \mbox{and}\label{eqn:slope}\\
    \mbox{coefficient of variation } c  & = & \frac{1}{\sqrt{\kappa}} \label{eqn:varcoef}.
  \end{eqnarray}
  From this, we can see that for small~$c$, the (logit-) toxicity
  behaves approximately linear, while larger $c$~values allow for
  departures from linearity. In the limiting case of linearity, the
  model simplifies to a logistic model, which, in the special case of
  dose increments defined on the logarithmic scale as suggested above,
  again is a special case of the \emph{Emax} model
  \citep{SchwinghammerKroboth1988}.

\subsection{Prior effective sample sizes for fixed effects}

  In order to assess how informative certain choices of priors and
  hyperprior parameters for the fixed effect are, the notion of the
  \emph{effective sample size (ESS)} can be used for the final
  calibration of the prior distributions and/or hyperprior parameters
  \citep{morita08}. In the present case, we suggest to compute the
  approximate ESS as follows: (i)~set the desired hyperparameters,
  (ii)~simulate from the resulting set of prior distributions,
  (iii)~for each simulated vector value, compute each $p_i$ using
  \eqref{eqn:eq_mean} without random effects, (iv)~approximate each
  $p_i$'s distribution by a $\betadistn(a_i,b_i)$, (v)~compute the
  approximate ESS as $\frac{1}{I} \sum_i(a_i+b_i)$, that is, the
  average of the ESS at each dose level.

\subsection{MTD estimation}
  
  A range of rules have been proposed for estimating MTDs; several
  examples are given in the following. The most popular way uses the
  posterior mean estimates of the parameters in \eqref{eqn:eq_mean}
  and selects the MTD as the dose whose estimated DLT probability is
  closest to the pre-specified target $\tau\in[0,1]$
  \citep{cheung11}. In the meta-analysis context, we may focus on the
  overall fixed effect; inverting from~\eqref{eqn:eq_mean}, we hence
  define
  \begin{equation} \label{eqn:piDef} \textstyle
    \pi_i = \logit^{-1} \Bigl( \sum_{j=1}^i \mu_j \Bigr)
  \end{equation} 
  where the inverse logit is given by $\logit^{-1}(x)=(1+\exp(-x))^{-1}$.
  From this, we may then derive
  \begin{equation}\label{eqn:mtdDef}
    \mbox{MTD} \; = \; d_j \mbox{,} 
    \qquad \mbox{where} \qquad
    j \;=\; \argmin_i \, \Bigl| \expect[\pi_i|y] - \tau \Bigr| \mbox{,}
  \end{equation}
  and where $\expect[\pi_i|y]$ denotes the posterior expectation of
  $\pi_i$.  The MTD is hence defined as the dose with estimated
  overall mean response closest to the targeted one.
  Alternatively, the posterior median may also be used instead of the
  mean in \eqref{eqn:mtdDef} \citep{ursino19}.
  
  In situations where investigators are particularly interested in
  overdose control, the classical \emph{escalation with overdose
    control (EWOC)} principle may also be applied, so that the
  MTD~$d_i$ is chosen as the largest dose satisfying
  \begin{equation}
    \prob \bigl(\pi_i \geq \tau \,\big|\, y\bigr) \;<\; \tau_o \mbox{,} 
    \label{eqn:EWOC}
  \end{equation}
  that is, the dose whose posterior probability of exceeding the
  toxicity threshold~$\tau$ is less than a pre-specified
  threshold~$\tau_o$ \citep{babb98,NeuenschwanderEtAl2015}. More
  complex rules, involving loss functions, such as the one applied for
  the Bayesian Logistic Regression Model, can be also used
  \citep{neuenschwander08}.

\section{Simulations}\label{sec:simul}

  We performed an extensive simulation study to evaluate the operating
  characteristics of the proposed method. The aim was to compare the
  percentages of correct MTD selection to the ones of the ZKO method
  in several scenarios.  A total of nine scenarios are proposed, with
  variations in the position of the MTD, the heterogeneity structure
  and/or the design of the simulated trial. Details are given in
  Section~\ref{sec:simulData}.  Then we performed a sensitivity analysis
  aiming at checking the impact of prior distribution/hyperparamter
  choices and of random-effects model misspecification; details are
  shown in Sections~\ref{sec:simulPriors} and~\ref{sec:simulSensitivity}.

\subsection{Data generation scenarios}\label{sec:simulData}

\begin{figure}
\includegraphics[width=\textwidth]{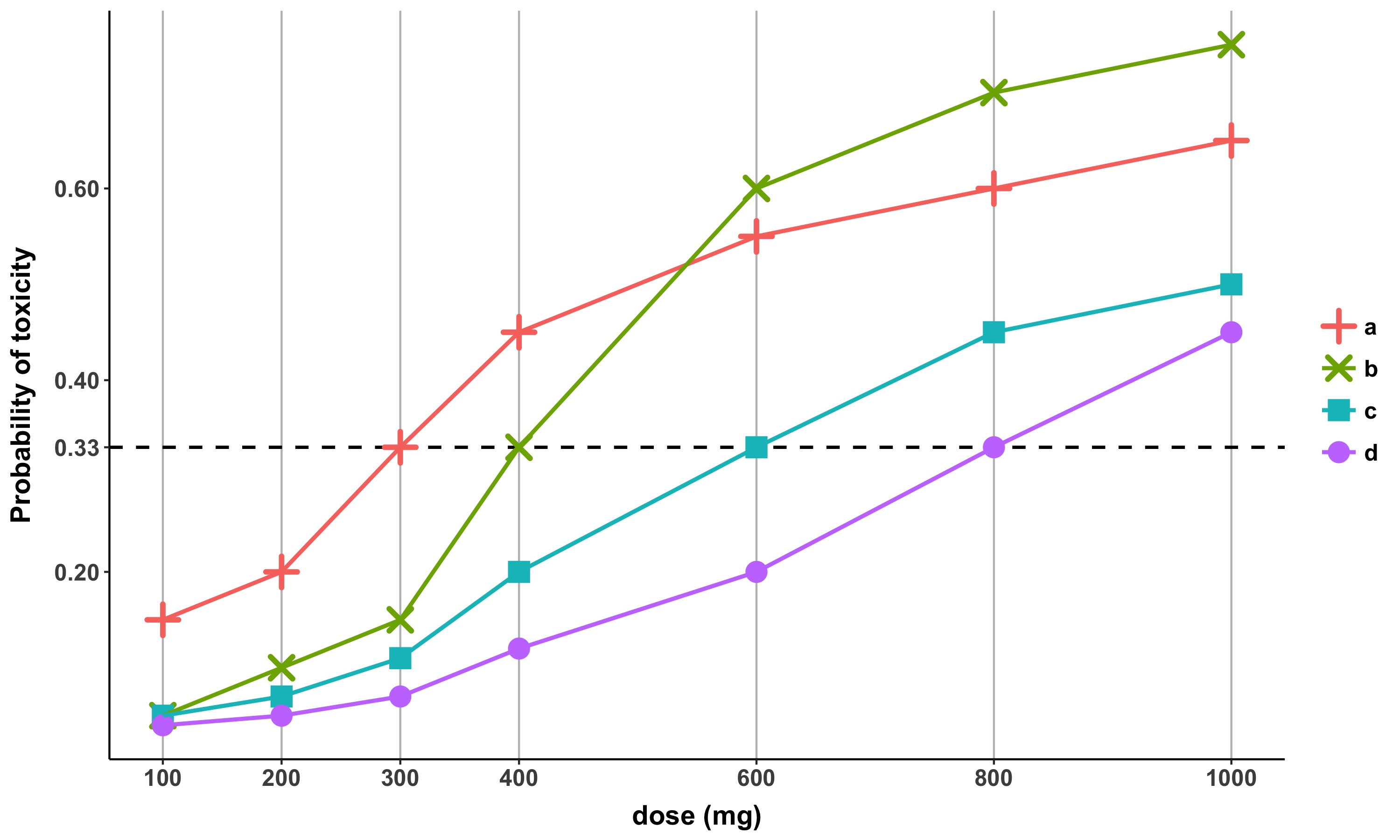}
\caption[]{Four different sets of probabilities~$\mathbf{p^\star}$ used to set the fixed effects in the data generation scenarios.
a: $\mathbf{p^\star}=$(0.15, 0.20, 0.33, 0,45, 0.55, 0.60, 0.65); b: $\mathbf{p^\star}=$(0.05, 0.10, 0.15, 0.33, 0.60, 0.70, 0.75); c: 
$\mathbf{p^\star}=$(0.05, 0.07, 0.11, 0.20, 0.33, 0.45, 0.50); d: $\mathbf{p^\star}=$(0.04, 0.05, 0.07, 0.12, 0.20, 0.33, 0.45).}
\label{fig:setfixeff}
\end{figure}

  For each scenario, we simulated 1000~sets of completed trials that
  were subsequently meta-analyzed.  Motivated by the sorafenib example
  (see Table~\ref{tab:sorafenib}), overall seven doses between
  $d_1=100$ mg and $d_7=1000$ mg were used.  We first set the true
  probabilities of toxicity of the scenario for each of the $I\!=\!7$
  doses involved, $\mathbf{p^\star} = (p^\star_1, \ldots, p^\star_I)$.
  Four different sets of $\mathbf{p^\star}$ were considered in total;
  these are illustrated in Figure~\ref{fig:setfixeff}.
  Then, the between-trial heterogeneity was added on the
  dose-transformed scale, in order to set the probabilities of
  toxicity used to generate each single trial.  However, since in our
  proposed model we used the logit transformation, in order to not
  generate data from the very same model, we opted for the probit
  function in data generation. Therefore, for the $k$th trial of the
  $j$th meta-analysis run, we first generated $\mathbf{p}_{kj}^{tr} =
  \mathcal{N}( \left( \Phi^{-1}(p_1^\star), \ldots ,
  \Phi^{-1}(p_I^\star) \right), \Sigma), $ where $\Phi(\cdot)$
  represents the cumulative distribution function of the standard
  normal distribution.  Then, we computed the probabilities as
  $\mathbf{p}_{kj} = \left( \Phi(p_{1,kj}^{tr}), \ldots ,
  \Phi(p_{I,kj}^{tr}) \right)$.  We used the same autocovariance
  structure as in the estimation model \eqref{eqn:OupCovariance} for
  all scenarios, allowing for a different $\sigmam$ value, except for
  scenario~6, where $\Sigma = \left[ \exp \left(- \frac{|\delta_{i,j}
      |}{\ell} \right) \sigma_i \sigma_j \right]$, and scenario~7,
  where $\Sigma = \left[ \exp \left(- \frac{\delta_{i,j}^2}{2\ell^2}
    \right) \sigma_m^2 \right]$.  For all scenarios, we set $\ell=1$
  and $\delta_{i,j}=\frac{d_i-d_j}{I^{-1} \sum_{n=1}^I d_i}$, while
  $\delta^*_{i,j}=\frac{d_i-d_j}{100\,\mathrm{mg}}$. This means that
  we used two related scales for $\delta$ and $\delta^*$, and that we
  utilise 100 mg as the measure unit for the fixed effect.
  
  The number of doses used for each trial is a random integer
  between~3 and~7 (sampled according to a uniform discrete
  distribution), and in all cases we have the true MTD (whose
  probability of toxicity equals the target of $\tau=0.33$) among the
  set of doses.  Then, complete patients' responses are drawn at each
  dose from a Binomial distribution \eqref{eqn:eventcount}.  Depending
  on the scenarios and on the total number of trials used in the
  meta-analysis, some of the trials followed a CRM design while others
  used the traditional ``3+3'' design\citep{OQuigley90, letourneau09}.
  For the CRM trials, the maximum sample size per study was sampled as
  an integer between 18 and 24 patients and the number of patients at
  each cohort between 2 and 3 (then, the maximum number of patients is
  automatically adjusted).
  
  The (estimated) MTD is defined as the dose whose probability of
  toxicity is closest to the target of $\tau=0.33$ and we adopted the
  posterior median variant of \eqref{eqn:mtdDef} as estimation rule.
  The skeleton, that is the prior guesses, was chosen to be (0.01,
  0.05, 0.1, 0.15, 0.25, 0.38, 0.45), where only the probabilities
  linked to the doses in the trial panel are used, and we selected the
  empirical working model.  Finally, the CRM trials adopted the
  ``no~skipping'' rule, that is, a higher dose is proposed to the next
  cohort only if all previous dose levels have already been given,
  while no stopping criteria were set.

  In Scenarios 1-4 the true MTD is shifted from dose level 3 to dose
  level 6, while keeping the same $\sigma = 0.3$. This allows us to
  test the impact of the number of doses and MTD position in the
  meta-analysis run. Scenarios 5 and 6 have the
  same~$\mathbf{p^\star}$ of Scenario 2, but we double the
  heterogeneity parameter in Scenario 5 and we allow for dose-specific
  heterogeneity in Scenario 6.  Then, Scenario 7 was added to check
  the impact of generating data under another Gaussian process.  We
  evaluated the performance of the proposed model in case of 10 trials
  (made by 5 CRM and 5 3+3) and 5 trials (3 CRM and 2 3+3) at each
  meta-analysis run.  In the last two scenarios, that is, Scenarios 8
  and 9, we evaluate the results given if all studies used an
  algorithm design (i.e. 3+3) or model based design (i.e. the CRM),
  respectively.  The simulation scenarios are summarised in
  Table~\ref{tab:scenarios}.

\begin{table}
\caption{Settings and parameters in the 9~different simulation scenarios.}
\label{tab:scenarios}
\begin{tabular}{ccll}
\hline
\multirow{2}{*}{Scenario} & Fixed effect  & \multicolumn{1}{c}{\multirow{2}{*}{Random effect}} & \multirow{2}{*}{Studies design} \\
& true $\mathbf{p^\star}$ &  &  \\
\hline
1 & a) & OUP, $\sigma = 0.3$ & CRM and 3+3 \\
2 & b) & OUP, $\sigma = 0.3$ & CRM and 3+3 \\
3 & c) & OUP, $\sigma = 0.3$ & CRM and 3+3 \\
4 & d) & OUP, $\sigma = 0.3$ & CRM and 3+3 \\
5 & b) & OUP, $\sigma = 0.6$ & CRM and 3+3  \\
\multirow{2}{*}{6} & b) & $\Sigma = \left[ \exp \left(- \frac{|\delta_{i,j} |}{l} \right) \sigma_i \sigma_j   \right]$  & \multirow{2}{*}{CRM and 3+3} \\
 & & and $\boldsymbol{\sigma} = (0.1, 0.1, 0.2, 0.3, 0.4, 0.5, 0.6)$  &\\
7 & c)  & $\Sigma = \left[ \exp \left(- \frac{\delta_{i,j}^2}{2l^2} \right) \sigma_m^2 \right]$, $\sigma = 0.3$ & CRM and 3+3  \\
8 & c) & OUP, $\sigma = 0.3$ & only 3+3 design \\
9 & c) & OUP, $\sigma = 0.3$ &  only CRM design \\
\hline
\end{tabular}
\end{table}

\subsection{Prior settings}\label{sec:simulPriors}

\begin{figure}
\includegraphics[width=\textwidth]{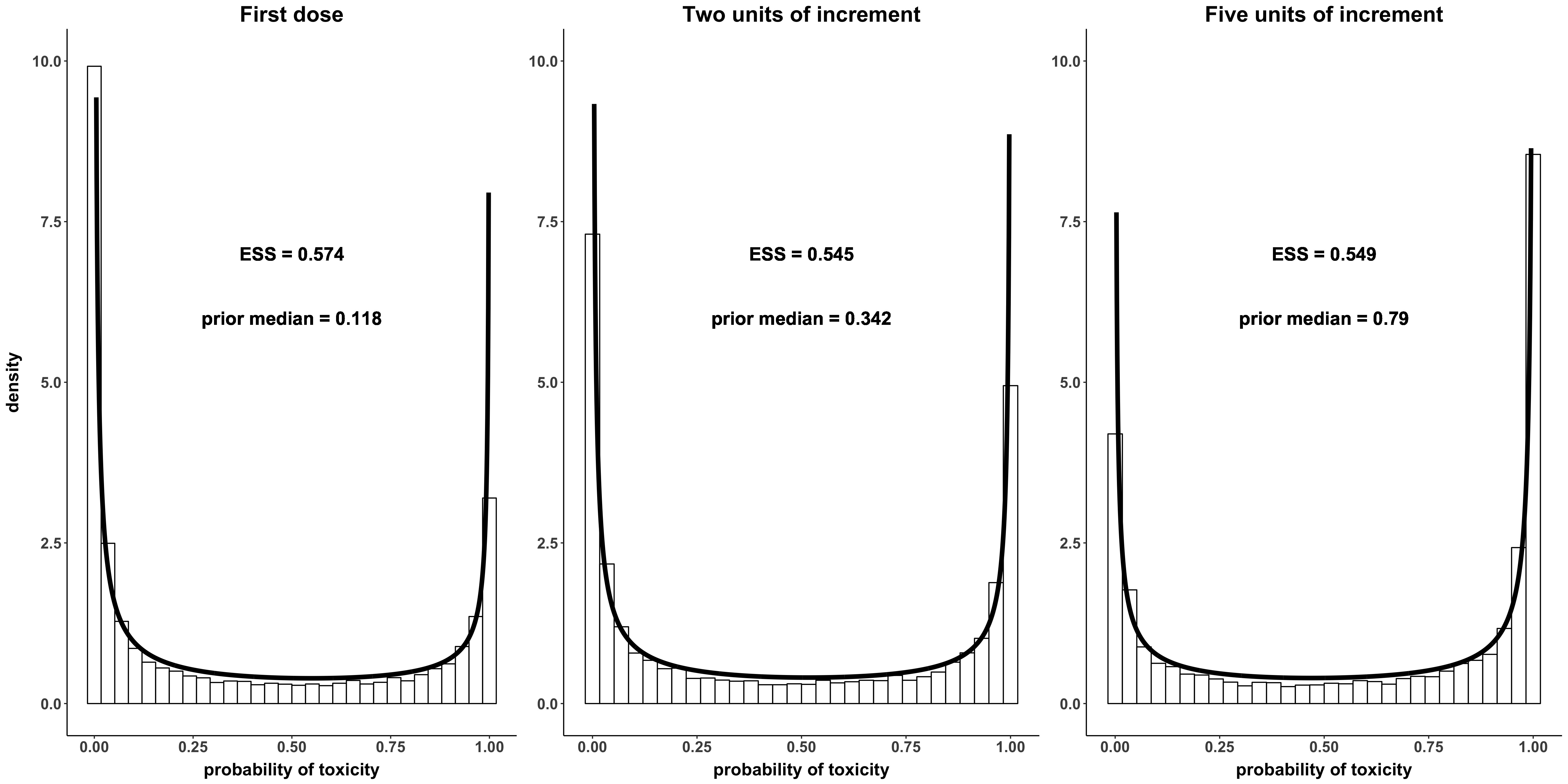}
\caption[]{Prior distribution shown by number of dose increments.}
\label{fig:prior01}
\end{figure}

\begin{figure}
\includegraphics[width=\textwidth]{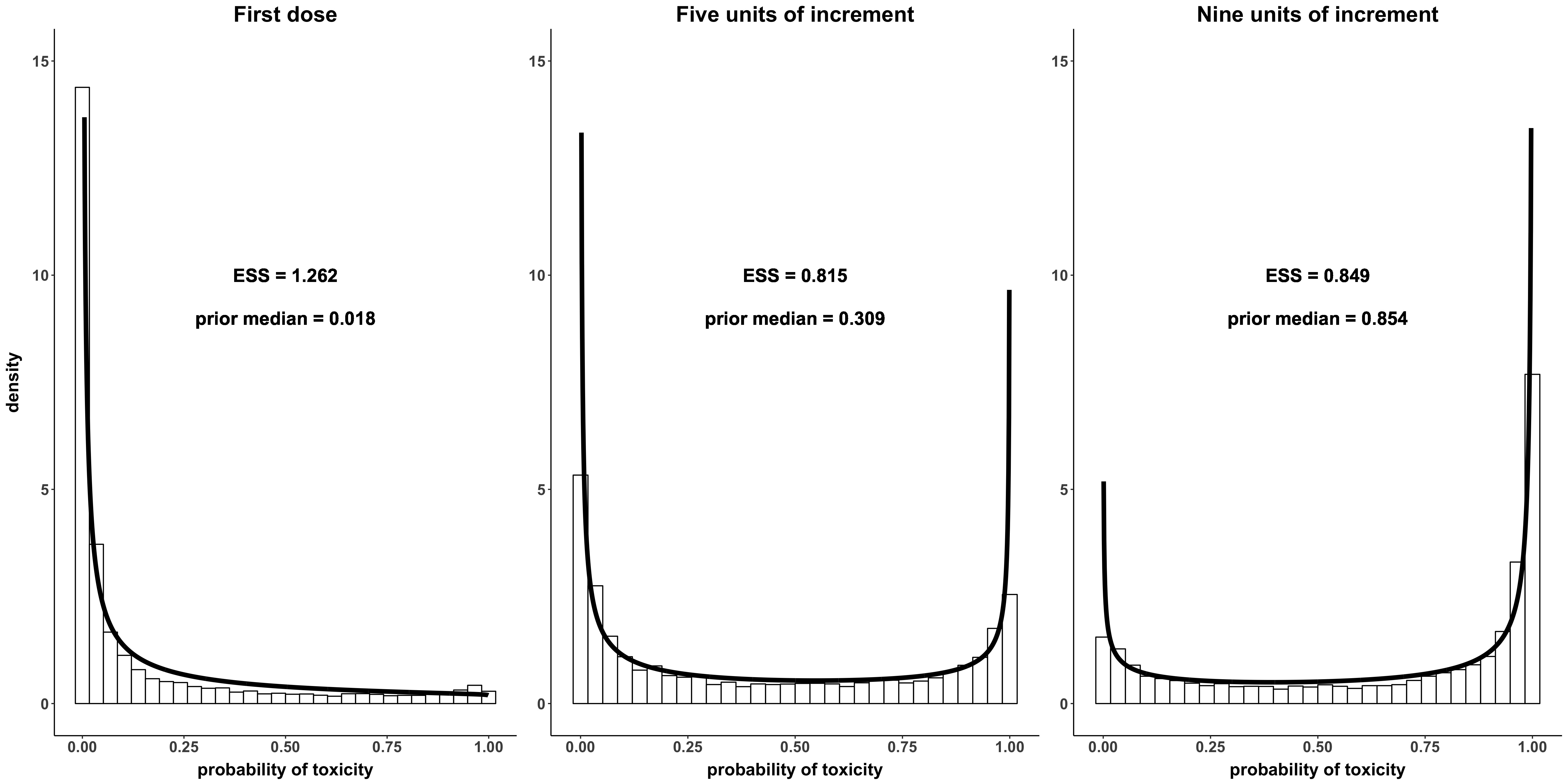}
\caption[]{Prior distribution shown by number of dose increments.}
\label{fig:prior02}
\end{figure}

  When running a single meta-analysis, the user knows in advance the
  number of doses in the analysis and it is natural to select prior
  distribution which suggest the MTD in the second half part of the
  dose panel. However, during simulations, depending on the scenarios,
  the number of doses in the panel and the related number of
  increments can vary considerably. Therefore, even if it is not
  strictly necessary in a single run, we used a variation of the
  empirical Bayes approach to adaptively select the prior parameters
  of the Gamma prior process, taking care about the number of dose
  increments in the actual run.  Specifically, we compute the
  empirical probability of toxicity of each dose by summing all DLTs
  reported on all studies at the same dose level and dividing it by
  the total number of patients treated at this dose level (in all
  studies). A linear order isotonic regression, which uses the pool
  adjacent violators algorithm, was then applied to assure the
  non-decreasing behaviour of the dose-toxicity curve.  Finally, the
  empirical MTD was selected as the dose whose empirical probability
  of toxicity is closest to the target, set as 0.33 in this simulation
  study.  The set of parameters was chosen looking at the difference
  between the selected MTD and the first dose in the panel: if the
  difference is less or equal to two units, we select $\mu^\ast=-2$,
  $\sigma^\ast=5$, $a=0.667$ and $c=0.5$ which gives the induced prior
  probability of toxicities shown in Figure~\ref{fig:prior01};
  otherwise, we select $\mu^\ast=-4$, $\sigma^\ast=3.5$, $a=0.642$ and
  $c=0.5$ which gives the induced prior probability of toxicities
  shown in Figure~\ref{fig:prior02}. These values were chosen in order
  to have a good trade-off between ESS (lower numbers are desirable to
  have weakly informative prior) and the prior MTD placed at second
  and fifth increment, respectively.
  
  Finally, a half-Normal distribution was chosen as prior distribution
  for $\sigmam$ and an inverse Gamma distribution with shape and scale
  equal to 1 for~$\ell$.
  
  The resulting model will be referred as MADF from now on.

\subsection{Sensitivity analyses}\label{sec:simulSensitivity}

  We performed sensitivity analyses to check the impact of prior
  distributions and/or random-effects model misspecification. We
  considered four model modifications, changing the prior distribution
  for the fixed effect, or changing the correlation structure for the
  random effects (or both).

  Let MADF1 denote the
  model MADF where \eqref{eqn:gamma_prior} is substituted by
  \[ \mu_i \;\sim\; \gammadistn (k=\kappa, \theta) \quad i>1,\] 
  that is, the process assumes identical dose increments and all
  $\mu_{i>1}$ have the same prior distribution.  In particular, we
  chose $\kappa = 3$ and $\theta=2$ which led to very ``pessimistic''
  prior probabilities of toxicities, that is, the prior probabilities
  of toxicities tends to be close to~1 for all doses larger than the
  first one.

  MADF2, instead, denotes the model MADF with $\Sigma$ as the
  variance-covariance of a heterogeneous first order autoregressive
  process, that is,
\[
\Sigma  = 
\begin{bmatrix}
    \sigma_1^2    & \rho \sigma_1 \sigma_2 & \rho^2 \sigma_1 \sigma_3 & \dots & \rho^{I-1} \sigma_1 \sigma_I \\
   \rho \sigma_1 \sigma_2     &\sigma_2^2 & \rho \sigma_2 \sigma_3 & \dots & \rho^{I-2} \sigma_2 \sigma_I \\
   \rho^2 \sigma_1 \sigma_3  &   \rho \sigma_2 \sigma_3 &\sigma_3^2 &  \dots & \rho^{I-3} \sigma_3 \sigma_I \\
    \hdotsfor{5} \\
    \rho^{I-1} \sigma_1 \sigma_I    & \dots  &  \dots & \dots & \sigma_I^2
\end{bmatrix},
\]
  along with a half-normal distribution with scale~1 as prior
  distribution for each $\sigma_i$, and a uniform distribution across
  the interval~$[0,1]$ for $\rho$.

  In MADF3, $\Sigma = \sigma \mathcal{I}$, where $\mathcal{I}_I$
  represents the identity matrix of $I \times I$ dimensions. In this
  case, random effects are uncorrelated and each dose has a proper
  scalar value. Again, $\sigma \sim \halfnormaldistn (0,1)$.

  Finally, MADF4 shares the same model of MADF3 except for the
  Gamma prior distribution, which is $\mu^\ast=-2$, $\sigma^\ast=7$,
  $a=3$ and $c=0.5$ if the increment is less or equal to two units;
  otherwise, $\mu^\ast=-4$, $\sigma^\ast=10$, $a=3$ and $c=2$.

\subsection{Results}

  Table~\ref{tab:results_1} shows the results in terms of percentage
  of correct MTD selection (PCS) of the proposed method, MADF, versus
  the ZKO, when 10 studies are included in each meta-analysis run.
  MADF has higher PCS, ranging from 0.61 to 0.92, while ZKO performs
  well in the range from 0.50 to 0.72. This could be expected, since
  ZKO does not take into account heterogeneity between trials. ZKO
  tends to select overdoses more often than MADF, for example in
  Scenario~1, where the MTD is at dose level~3, MADF suggests 34\%
  over toxic doses versus 41\% of ZKO\@. PCS percentages decrease as
  $\sigma$ increases, as in Scenario~5, and are stable for
  random-effects misspecification, as in Scenarios~6 and~7.

  These percentages decrease when only 5~studies are incorporated in
  the meta-analysis. The results are shown in
  Table~\ref{tab:results_2}, where MADF has still higher PCS, ranging
  from 0.5 to 0.82, while ZKO has performance ranging from 0.38 to
  0.61.

\begin{table}
\caption{Percentage of dose selection with 10 studies for each meta-analysis method.}
\label{tab:results_1}
\setlength\tabcolsep{3pt} % default value: 6pt
\scriptsize{\begin{tabular}{l c c c c c c c }
\hline
 & \multicolumn{7}{c}{Dose levels} \\
  & 1 & 2 & 3 & 4 & 5 & 6 & 7   \\
\hline
Scenario 1 & \multicolumn{7}{l}{}  \\
MADF & 0.000 & 0.082 & \textbf{0.612} & 0.305 & 0.001 & 0.000 & 0.000 \\ 
  ZKO & 0.022 & 0.190 & \textbf{0.496} & 0.253 & 0.034 & 0.002 & 0.003 \\ 
  \#patients & 31 (23, 41) & 31 (23, 41) & 54 (43, 65) & 15 (9, 23) & 6 (3, 12) & 2 (0, 6) & 0 (0, 3) \\ 
 \multicolumn{8}{l}{}  \\  
 Scenario 2 & \multicolumn{7}{l}{}  \\
 MADF & 0.000 & 0.000 & 0.032 & \textbf{0.920} & 0.048 & 0.000 & 0.000 \\ 
  ZKO & 0.000 & 0.002 & 0.052 & \textbf{0.695} & 0.233 & 0.013 & 0.005 \\ 
  \#patients & 22 (18, 26) & 26 (20, 32) & 29 (23, 37) & 59 (50, 68) & 14 (9, 20) & 5 (0, 9) & 0 (0, 3) \\ 
   \multicolumn{8}{l}{}  \\  
 Scenario 3 & \multicolumn{7}{l}{}  \\
 MADF & 0.000 & 0.000 & 0.000 & 0.084 & \textbf{0.834} & 0.082 & 0.000 \\ 
  ZKO & 0.000 & 0.000 & 0.002 & 0.075 & \textbf{0.676} & 0.216 & 0.031 \\ 
  \#patients & 22 (17, 26) & 23 (19, 29) & 26 (20, 33) & 29 (22, 38) & 45 (36, 54) & 11.5 (6, 18) & 6 (2, 12) \\ 
     \multicolumn{8}{l}{}  \\  
 Scenario 4 & \multicolumn{7}{l}{}  \\
 MADF & 0.000 & 0.000 & 0.000 & 0.000 & 0.228 & \textbf{0.758} & 0.014 \\ 
  ZKO & 0.000 & 0.000 & 0.000 & 0.001 & 0.131 & \textbf{0.680} & 0.188 \\ 
  \#patients & 43 (37, 51) & 43 (37, 51) & 24 (19, 31) & 26 (20, 34) & 26 (20, 33) & 40 (32, 48) & 11 (6, 18) \\ 
    \multicolumn{8}{l}{}  \\  
 Scenario 5 & \multicolumn{7}{l}{}  \\
 MADF & 0.000 & 0.000 & 0.085 & \textbf{0.781} & 0.134 & 0.000 & 0.000 \\ 
  ZKO & 0.004 & 0.037 & 0.162 & \textbf{0.561} & 0.215 & 0.017 & 0.004 \\ 
  \#patients & 24 (19, 31) & 27 (20, 35) & 28 (21, 37) & 51 (41, 59) & 13 (8, 20) & 6 (2, 12) & 0 (0, 6) \\ 
    \multicolumn{8}{l}{}  \\  
 Scenario 6 & \multicolumn{7}{l}{}  \\
 MADF & 0.000 & 0.000 & 0.019 & \textbf{0.882} & 0.099 & 0.000 & 0.000 \\ 
  ZKO & 0.000 & 0.000 & 0.022 & \textbf{0.665} & 0.287 & 0.015 & 0.011 \\ 
  \#patients & 21 (17, 26) & 25 (20, 30) & 30 (23, 37) & 61 (53, 69) & 14 (8, 20) & 5 (0, 9) & 0 (0, 4) \\ 
    \multicolumn{8}{l}{}  \\  
 Scenario 7 & \multicolumn{7}{l}{}  \\
 MADF & 0.000 & 0.000 & 0.000 & 0.069 & \textbf{0.830} & 0.101 & 0.000 \\ 
  ZKO & 0.000 & 0.000 & 0.002 & 0.075 & \textbf{0.653} & 0.245 & 0.025 \\ 
  \#patients & 22 (18, 26) & 22.5 (18, 28) & 27 (20, 33.25) & 30 (23, 38) & 45 (36, 54) & 12 (6, 18) & 6 (3, 12) \\
     \multicolumn{8}{l}{}  \\  
 Scenario 8 & \multicolumn{7}{l}{}  \\
 MADF & 0.000 & 0.000 & 0.000 & 0.150 & \textbf{0.773} & 0.077 & 0.000 \\ 
  ZKO & 0.000 & 0.000 & 0.002 & 0.078 & \textbf{0.591} & 0.295 & 0.034 \\ 
  \#patients & 24 (18, 27) & 24 (18, 27) & 24 (18, 27) & 24 (18, 27) & 30 (24, 36) & 9 (3, 12) & 3 (0, 6) \\ 
     \multicolumn{8}{l}{}  \\  
 Scenario 9 & \multicolumn{7}{l}{}  \\
 MADF & 0.000 & 0.000 & 0.000 & 0.064 & \textbf{0.837} & 0.099 & 0.000 \\ 
  ZKO & 0.000 & 0.000 & 0.001 & 0.076 & \textbf{0.715} & 0.194 & 0.014 \\ 
  \#patients & 20 (16, 25) & 24 (18, 31) & 30 (23, 39) & 37 (27, 47) & 60 (49, 71) & 15 (9, 23) & 10 (4, 16) \\ 
\hline
\end{tabular}}
\end{table}

\begin{table}
\caption{Percentage of dose selection with 5 studies for each meta-analysis method.}
\label{tab:results_2}
\setlength\tabcolsep{3pt} % default value: 6pt
\scriptsize{\begin{tabular}{l c c c c c c c }
\hline
 & \multicolumn{7}{c}{Dose levels} \\
  & 1 & 2 & 3 & 4 & 5 & 6 & 7   \\
\hline
Scenario 1 & \multicolumn{7}{l}{}  \\
MADF & 0.007 & 0.153 & \textbf{0.498} & 0.324 & 0.018 & 0.000 & 0.000 \\ 
  ZKO & 0.032 & 0.177 & \textbf{0.377} & 0.292 & 0.089 & 0.026 & 0.007 \\ 
  \#patients & 15 (10, 23) & 16 (10, 23) & 29 (21, 37) & 8 (3, 14) & 3 (0, 6) & 0 (0, 3) & 0 (0, 0) \\ 
 \multicolumn{8}{l}{}  \\  
 Scenario 2 & \multicolumn{7}{l}{}  \\
MADF & 0.000 & 0.002 & 0.068 & \textbf{0.826} & 0.103 & 0.001 & 0.000 \\ 
  ZKO & 0.005 & 0.007 & 0.060 & \textbf{0.490} & 0.367 & 0.055 & 0.016 \\ 
  \#patients & 11 (8, 14) & 12 (9, 17) & 15 (9, 21) & 32 (26, 38) & 6 (3, 12) & 0 (0, 6) & 0 (0, 0) \\ 
   \multicolumn{8}{l}{}  \\  
 Scenario 3 & \multicolumn{7}{l}{}  \\
MADF & 0.000 & 0.000 & 0.003 & 0.169 & \textbf{0.683} & 0.144 & 0.001 \\ 
  ZKO & 0.000 & 0.000 & 0.013 & 0.133 & \textbf{0.544} & 0.261 & 0.049 \\ 
  \#patients & 11 (8, 14) & 11 (8, 15) & 13 (9, 18) & 15 (9, 21) & 24 (18, 30) & 6 (2, 10) & 2 (0, 6) \\ 
     \multicolumn{8}{l}{}  \\  
 Scenario 4 & \multicolumn{7}{l}{}  \\
MADF & 0.000 & 0.000 & 0.000 & 0.003 & 0.320 & \textbf{0.622} & 0.055 \\ 
  ZKO & 0.000 & 0.000 & 0.001 & 0.013 & 0.179 & \textbf{0.610} & 0.197 \\ 
  \#patients & 21 (17, 26) & 21 (17, 26) & 12 (8, 16) & 12 (8, 18) & 12.5 (9, 18) & 20 (14, 27) & 6 (0, 11) \\ 
    \multicolumn{8}{l}{}  \\  
 Scenario 5 & \multicolumn{7}{l}{}  \\
MADF & 0.000 & 0.017 & 0.153 & \textbf{0.622} & 0.200 & 0.008 & 0.000 \\ 
  ZKO & 0.015 & 0.045 & 0.117 & \textbf{0.436} & 0.299 & 0.076 & 0.012 \\
  \#patients & 11 (8, 16) & 13 (9, 19) & 14 (9, 20) & 27 (20, 34) & 6 (3, 12) & 3 (0, 6) & 0 (0, 3) \\ 
    \multicolumn{8}{l}{}  \\  
 Scenario 6 & \multicolumn{7}{l}{}  \\
MADF & 0.000 & 0.000 & 0.059 & \textbf{0.802} & 0.137 & 0.002 & 0.000 \\ 
  ZKO & 0.000 & 0.002 & 0.04 & \textbf{0.449} & 0.412 & 0.065 & 0.032 \\ 
  \#patients & 11 (8, 14) & 12 (9, 16) & 15 (10, 21) & 33 (27, 39) & 6 (3, 12) & 2 (0, 6) & 0 (0, 2) \\
    \multicolumn{8}{l}{}  \\  
 Scenario 7 & \multicolumn{7}{l}{}  \\
MADF & 0.000 & 0.000 & 0.001 & 0.152 & \textbf{0.692} & 0.155 & 0.000 \\ 
  ZKO & 0.001 & 0.000 & 0.007 & 0.106 & \textbf{0.546} & 0.271 & 0.069 \\
  \#patients & 11 (8, 14) & 11 (8, 15) & 13 (9, 18) & 15 (9, 21) & 24 (17, 30) & 6 (2, 11) & 3 (0, 6) \\ 
     \multicolumn{8}{l}{}  \\  
 Scenario 8 & \multicolumn{7}{l}{}  \\
MADF & 0.000 & 0.001 & 0.009 & 0.203 & \textbf{0.668} & 0.116 & 0.003 \\ 
  ZKO &  0.000 & 0.001 & 0.017 & 0.134 & \textbf{0.489} & 0.282 & 0.077 \\ 
  \#patients & 12 (9, 15) & 12 (9, 15) & 12 (9, 15) & 12 (9, 15) & 15 (12, 18) & 3 (0, 6) & 0 (0, 3) \\
     \multicolumn{8}{l}{}  \\  
 Scenario 9 & \multicolumn{7}{l}{}  \\
MADF & 0.000 & 0.000 & 0.002 & 0.143 & \textbf{0.713} & 0.141 & 0.001 \\ 
  ZKO & 0.000 & 0.000 & 0.005 & 0.125 & \textbf{0.557} & 0.269 & 0.044 \\ 
  \#patients & 10 (7, 13) & 11 (8, 16) & 14 (9, 21) & 18 (11.75, 25) & 30 (22, 38) & 7 (2, 13) & 3 (0, 10) \\ 
\hline
\end{tabular}}
\end{table}

\begin{figure}
\includegraphics[width=\textwidth]{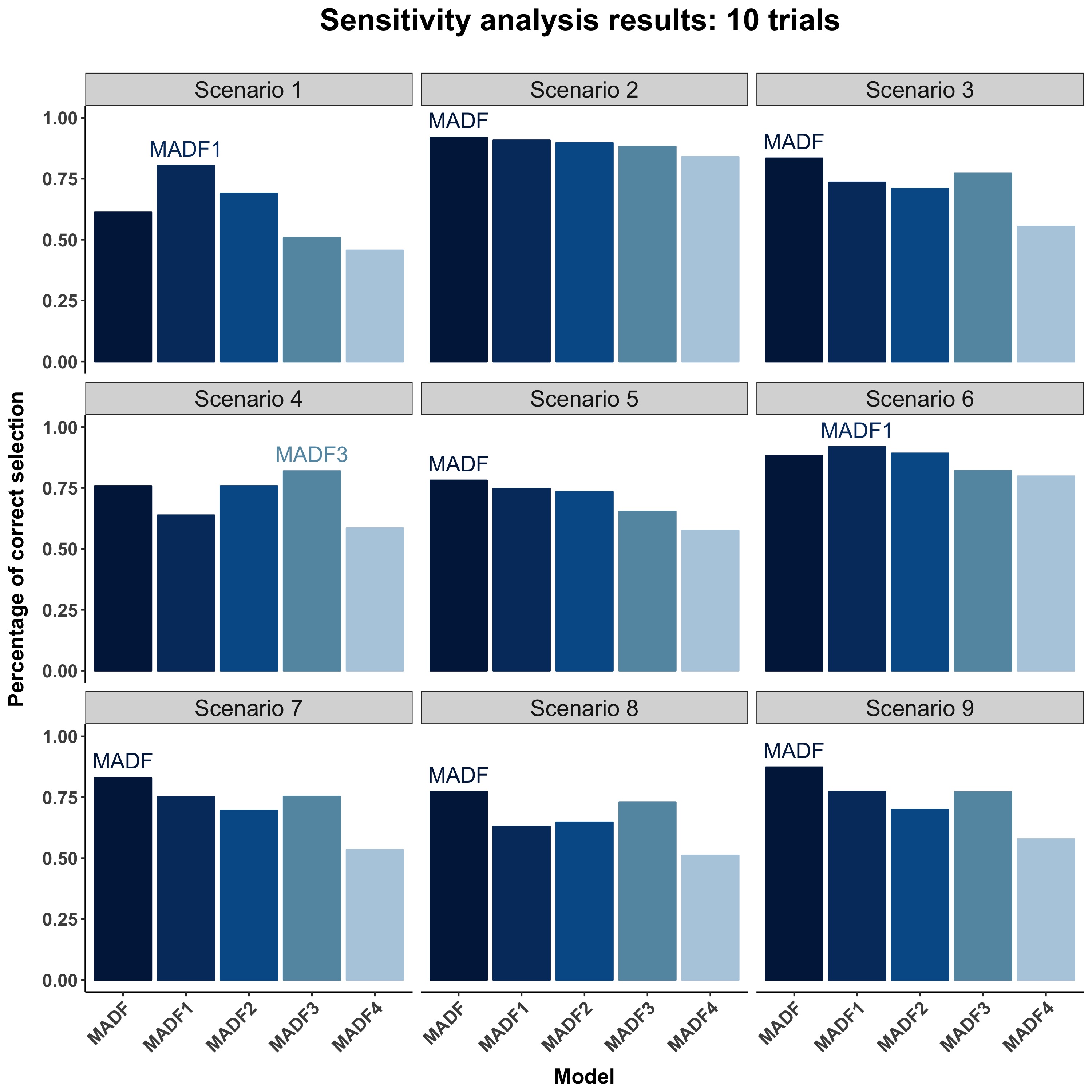}
\caption[]{Results in terms of percentage of correct selection when
  10 studies are adopted in each analysis. In each scenario, the name of the method with the high percentage of correct selection is shown.}
\label{fig:res_sensplot}
\end{figure}

  Figure~\ref{fig:res_sensplot} resumes the results of the sensitivity
  analysis in terms of percentage of correct selection when 10 studies
  are adopted in each analysis. MADF1 has the best performance in
  Scenarios 1 and 6, while MADF3 is the best method in Scenario
  4. MADF4 has the lowest PCS in all scenarios. Full results are given
  in Table~\ref{tab:results_sens1} in the Appendix.  We can see the
  same trend for 5 studies, except in Scenario 4, where MADF1 gets the
  lowest PCS (full results given in Table~\ref{tab:results_sens2} in
  Appendix).

\section{Application to case studies}\label{sec:application}
\subsection{The sorafenib example}

\begin{figure} 
\includegraphics[width=0.9\textwidth]{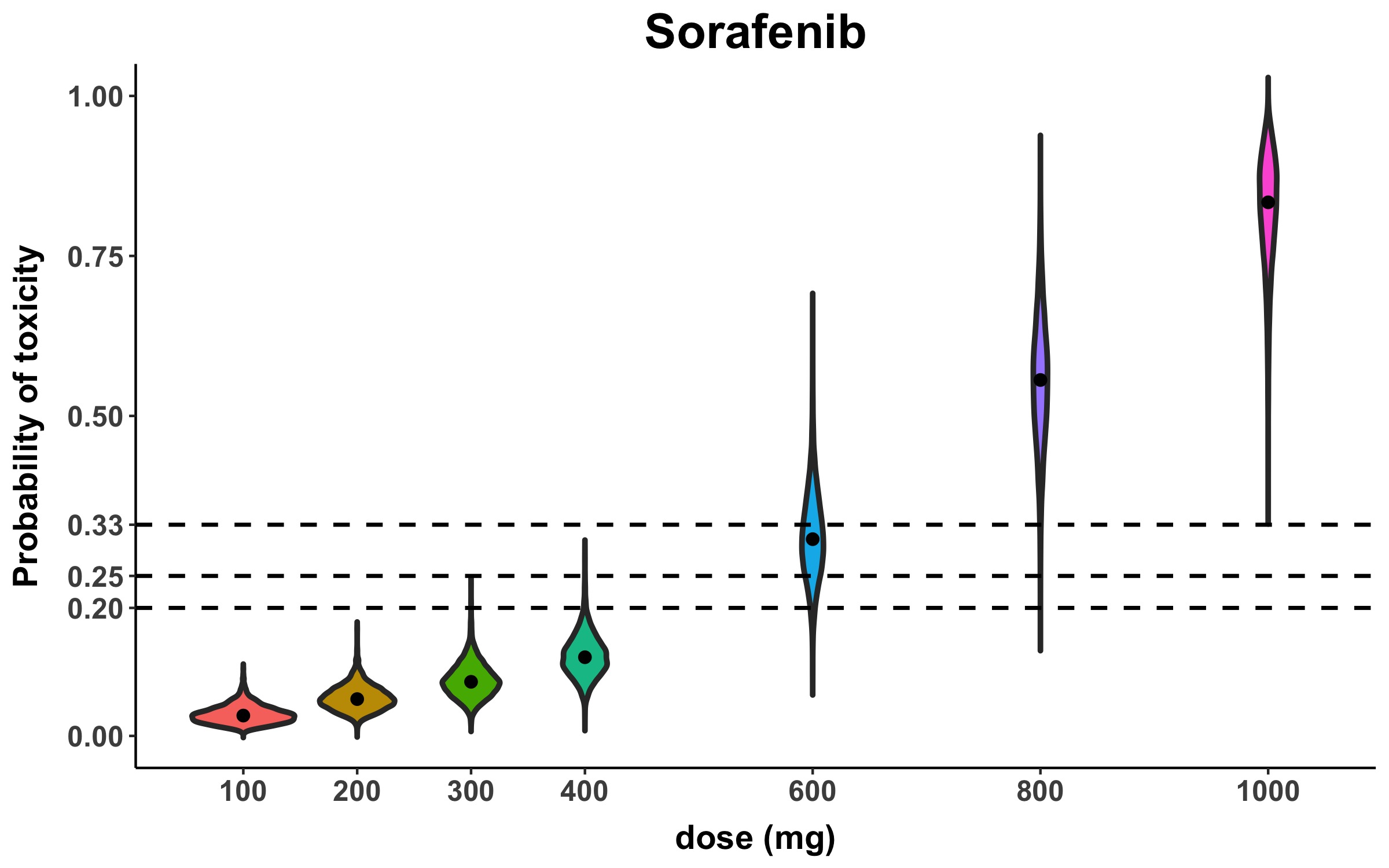}
\caption[]{Posterior distributions for dose-limiting toxicity probability each dose level
  for the sorafenib example.}
\label{fig:res_sorafenib}
\end{figure}

  We applied the MADF method, with the same setting and prior
  distributions as described in the previous section, to both examples
  introduced in Section~\ref{sec:motivation}.  Regarding the sorafenib
  example, Figure~\ref{fig:res_sorafenib} shows the posterior
  distribution obtained for the probability of toxicity associated to
  each dose panel level.  Using the posterior median variant of
  \eqref{eqn:mtdDef}, we obtain the following estimates (0.032, 0.058,
  0.085, 0.123, 0.307, 0.556, 0.834). This leads to selecting dose 600
  mg as MTD if $\tau=0.33$ or $\tau=0.25$, while 400 mg is chosen when
  $\tau=0.20$. Adopting the EWOC rules as in \eqref{eqn:EWOC}, that is
  computing $\prob \bigl(\pi_i \geq \tau \,\big|\, y\bigr)$, we obtain
  (0, 0, 0, 0, 0.369, 0.991, 1), (0, 0, 0, 0.002, 0.832, 1, 1) and (0,
  0, 0.001, 0.016, 0.964, 1, 1) for $\tau=0.33$, $\tau=0.25$ and
  $\tau=0.20$, respectively. Setting $\tau_o = 0.25$, we select dose
  400 mg in all cases.

\subsection{The irinotecan / S-1 example}

\begin{figure} 
\includegraphics[width=0.9\textwidth]{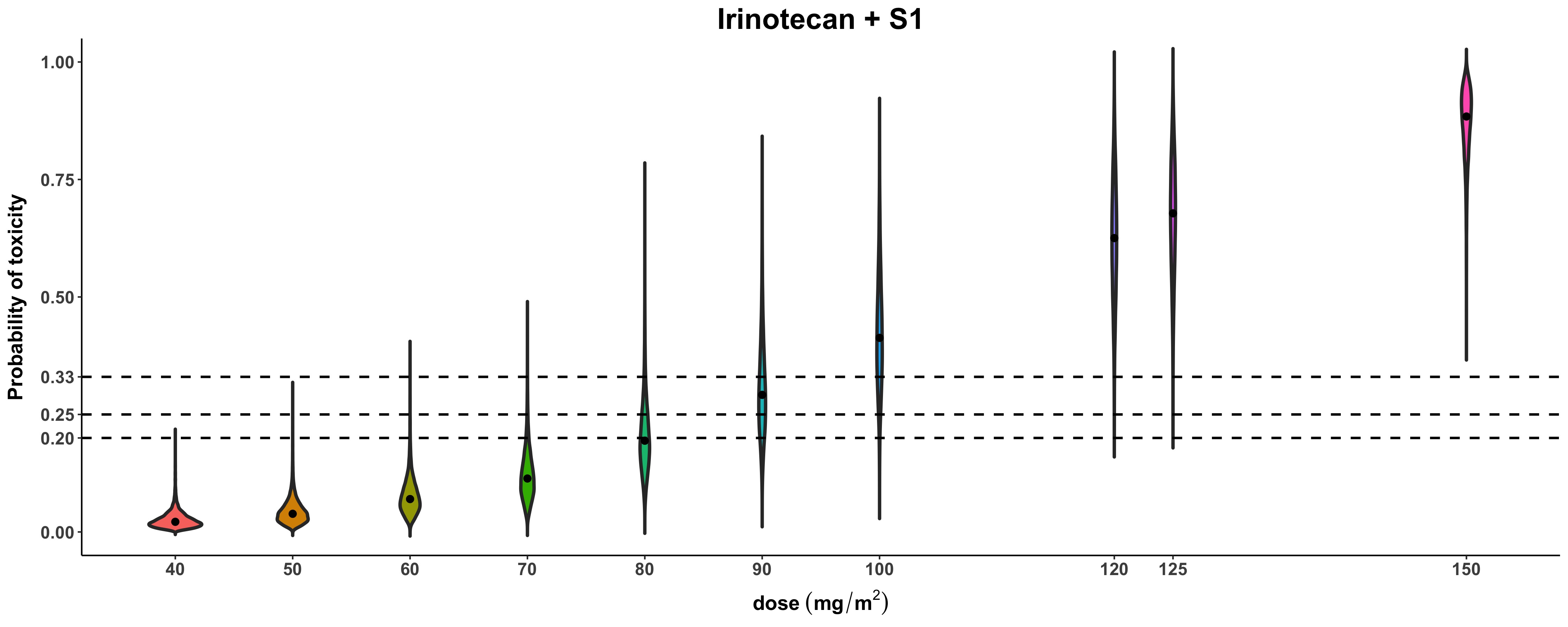}
\caption[]{Posterior distributions for dose-limiting toxicity probability at each dose level for the irinotecan + S-1 example.}
\label{fig:res_irinotecan}
\end{figure}

  Results of the irinotecan + S-1 example are shown in
  Figure~\ref{fig:res_irinotecan}. Differently from before, here
  $\delta^*_{i,j}=\frac{d_i-d_j}{10\,\mathrm{mg}\,\mathrm{m}^{-2}}$,
  while $\delta$ has the same specification as before.  Again, using
  the posterior median variant of \eqref{eqn:mtdDef}, we obtain the
  following estimates (0.022, 0.039, 0.070, 0.114, 0.194, 0.292,
  0.413, 0.625, 0.678, 0.884). This leads to selecting dose 90
  mg/m$^2$ as MTD if $\tau=0.33$ or $\tau=0.25$, while 80 mg/m$^2$ is
  chosen when $\tau=0.20$. Adopting the EWOC rules as in
  \eqref{eqn:EWOC}, we obtain (0, 0, 0, 0.004, 0.061, 0.349, 0.773,
  0.990, 0.996, 1), (0, 0, 0.002, 0.027, 0.238, 0.677, 0.944, 0.998,
  1, 1) and (0, 0.001, 0.008, 0.082, 0.466, 0.866, 0.984, 1, 1, 1) for
  $\tau=0.33$, $\tau=0.25$ and $\tau=0.20$, respectively. Setting
  $\tau_o = 0.25$, we select dose 80 mg/m$^2$ in the first two cases
  and 70 mg/m$^2$ in the last one.

\section{Discussion}\label{sec:discussion}

  We proposed a new methodology for random-effects meta-analysis of
  phase~I dose-finding trials, based on a Gaussian process for the
  random effect structure, and a Gamma process as a prior distribution
  for the fixed effects. The Gaussian process permits to share more
  information when doses are closer and less information when they are
  distant. In this way, for example, regarding a dose panel, dose
  level 3 and 4 are more correlated than dose level 1 and 4. And the
  amount of correlation depends on the distance, that seems more
  logical than assuming a constant value for the correlation. The
  Gamma prior process preserves the monotonicity assumption of
  toxicity. We do not suggest to add the full process to be estimated,
  since, in our experience, even if in meta-analysis more data are
  available than a single dose-finding trial, data are still not
  sufficient for a good estimation of the process parameters (results
  not shown in the paper). To note, we focused on modelling toxicities
  exactly at the doses~$d_i$ that had also been investigated in the
  analysed trials. In general, in case of rich data and when the
  estimation of the underlying Gamma process is feasible, the full
  model actually also allows to interpolate or extrapolate across the
  continuum of doses. In this case, guidance on how to set the prior
  distributions can be found in \citet{GelmanEtAl2008}.
  
  Since the two metrics used in the fixed effect prior and random
  effect, that is $\delta^*$ and $\delta$, respectively, are linked to
  each others via linear transformation, one can also consider to use
  the same metric and scaling the prior distributions accordingly.
  
  With the above model specifications, we have generalized the BNHM,
  as an obvious approach for the single-dose case, to the case of
  several adjacent doses. Note that for the special case of a single
  dose, we actually again recover the BNHM with the parameters~$\mu_1$
  and~$\sigmam$ corresponding to the overall mean and heterogeneity
  parameters.
 
  In our results, ZKO had lower PCS performance. This is expected,
  since this method does not take into account the heterogeneity
  between trials. Also as expected, PCSs decrease when heterogeneity
  increases and when only 3+3 dose-finding trials are incorporated in
  meta-analysis (scenario 5 and 8, respectively).  MADF showed to be
  stable to models misspecification, as we can see in the results of
  scenario 6 and 7 compared to scenario 2 and 3, respectively. On the
  other hand, prior specification and simpler models, as MADF3 and
  MADF4 can give different operating characteristic. A conservative
  Gamma process prior, as MADF1, has better PCS when the MTD is
  located at the beginning of the dose panel. Actually, this situation
  is not very realistic, since it would imply that really few safe
  doses where repeatedly tested in several clinical trials.

\appendix
\section*{Appendix}
\subsection*{Sensitivity analysis tables}\label{sec:tables}

  Tables~\ref{tab:results_sens1} and~\ref{tab:results_sens2} show the
  full results, in terms of percentage of MTD selection, of the
  sensitivity analysis performed.
  
\begin{table}
\caption{Sensitivity analysis results: percentage of correct selection with 10 studies for each meta-analysis method.}
\label{tab:results_sens1}
\setlength\tabcolsep{8pt} % default value: 6pt
\scriptsize{\begin{tabular}{l c c c c c c c }
\hline
 & \multicolumn{7}{c}{Dose levels} \\
  & 1 & 2 & 3 & 4 & 5 & 6 & 7   \\
\hline
Scenario 1 & \multicolumn{7}{l}{}  \\
MADF1 & 0.000 & 0.118 & \textbf{0.804} & 0.078 & 0.000 & 0.000 & 0.000 \\ 
  MADF2 & 0.000 & 0.071 & \textbf{0.690} & 0.232 & 0.007 & 0.000 & 0.000 \\ 
  MADF3 & 0.000 & 0.049 & \textbf{0.508} & 0.422 & 0.021 & 0.000 & 0.000 \\ 
  MADF4 & 0.000 & 0.045 & \textbf{0.456} & 0.406 & 0.088 & 0.004 & 0.001 \\ 
 \multicolumn{8}{l}{}  \\  
 Scenario 2 & \multicolumn{7}{l}{}  \\
MADF1 & 0.000 & 0.000 & 0.079 & \textbf{0.908} & 0.013 & 0.000 & 0.000 \\ 
  MADF2 & 0.000 & 0.000 & 0.043 & \textbf{0.897} & 0.060 & 0.000 & 0.000 \\ 
  MADF3 & 0.000 & 0.000 & 0.017 & \textbf{0.882} & 0.101 & 0.000 & 0.000 \\ 
  MADF4 & 0.000 & 0.000 & 0.026 & \textbf{0.840} & 0.131 & 0.003 & 0.000 \\
   \multicolumn{8}{l}{}  \\  
 Scenario 3 & \multicolumn{7}{l}{}  \\
MADF1 & 0.000 & 0.000 & 0.000 & 0.229 & \textbf{0.735} & 0.036 & 0.000 \\ 
  MADF2 & 0.000 & 0.000 & 0.000 & 0.101 & \textbf{0.709} & 0.185 & 0.005 \\ 
  MADF3 & 0.000 & 0.000 & 0.000 & 0.056 & \textbf{0.773} & 0.171 & 0.000 \\ 
  MADF4 & 0.000 & 0.000 & 0.000 & 0.061 & \textbf{0.554} & 0.329 & 0.056 \\ 
     \multicolumn{8}{l}{}  \\  
 Scenario 4 & \multicolumn{7}{l}{}  \\
MADF1 & 0.000 & 0.000 & 0.000 & 0.002 & 0.354 & \textbf{0.638} & 0.006 \\ 
  MADF2 & 0.000 & 0.000 & 0.000 & 0.000 & 0.132 & \textbf{0.758} & 0.110 \\ 
  MADF3 & 0.000 & 0.000 & 0.000 & 0.000 & 0.139 & \textbf{0.819} & 0.042 \\ 
  MADF4 & 0.000 & 0.000 & 0.000 & 0.000 & 0.078 & \textbf{0.585} & 0.337 \\ 
   \multicolumn{8}{l}{}  \\  
 Scenario 5 & \multicolumn{7}{l}{}  \\
MADF1 & 0.000 & 0.001 & 0.216 & \textbf{0.747} & 0.036 & 0.000 & 0.000 \\ 
  MADF2 & 0.000 & 0.001 & 0.105 & \textbf{0.734} & 0.156 & 0.004 & 0.000 \\ 
  MADF3 & 0.000 & 0.000 & 0.043 & \textbf{0.653} & 0.303 & 0.001 & 0.000 \\ 
  MADF4 & 0.000 & 0.000 & 0.051 & \textbf{0.575} & 0.337 & 0.037 & 0.000 \\
      \multicolumn{8}{l}{}  \\  
 Scenario 6 & \multicolumn{7}{l}{}  \\
MADF1 & 0.000 & 0.000 & 0.053 & \textbf{0.918} & 0.029 & 0.000 & 0.000 \\ 
  MADF2 & 0.000 & 0.000 & 0.028 & \textbf{0.892} & 0.080 & 0.000 & 0.000 \\ 
  MADF3 & 0.000 & 0.000 & 0.009 & \textbf{0.820} & 0.171 & 0.000 & 0.000 \\ 
  MADF4 & 0.000 & 0.000 & 0.022 & \textbf{0.798} & 0.172 & 0.008 & 0.000 \\ 
      \multicolumn{8}{l}{}  \\  
 Scenario 7 & \multicolumn{7}{l}{}  \\
MADF1 & 0.000 & 0.000 & 0.000 & 0.208 & \textbf{0.751} & 0.041 & 0.000 \\ 
  MADF2 & 0.000 & 0.000 & 0.001 & 0.088 & \textbf{0.696} & 0.210 & 0.005 \\ 
  MADF3 & 0.000 & 0.000 & 0.000 & 0.049 & \textbf{0.753} & 0.198 & 0.000 \\ 
  MADF4 & 0.000 & 0.000 & 0.000 & 0.050 & \textbf{0.534} & 0.349 & 0.067 \\ 
     \multicolumn{8}{l}{}  \\  
 Scenario 8 & \multicolumn{7}{l}{}  \\
MADF1 & 0.000 & 0.000 & 0.003 & 0.338 & \textbf{0.630} & 0.029 & 0.000 \\ 
  MADF2 & 0.000 & 0.000 & 0.002 & 0.149 & \textbf{0.647} & 0.192 & 0.010 \\ 
  MADF3 & 0.000 & 0.000 & 0.000 & 0.085 & \textbf{0.730} & 0.185 & 0.000 \\ 
  MADF4 & 0.000 & 0.000 & 0.001 & 0.077 & \textbf{0.511} & 0.339 & 0.072 \\ 
     \multicolumn{8}{l}{}  \\  
 Scenario 9 & \multicolumn{7}{l}{}  \\
MADF1 & 0.000 & 0.000 & 0.001 & 0.167 & \textbf{0.773} & 0.059 & 0.000 \\ 
  MADF2 & 0.000 & 0.000 & 0.000 & 0.091 & \textbf{0.699} & 0.205 & 0.005 \\ 
  MADF3 & 0.000 & 0.000 & 0.000 & 0.044 & \textbf{0.771} & 0.185 & 0.000 \\ 
  MADF4 & 0.000 & 0.000 & 0.000 & 0.048 & \textbf{0.578} & 0.331 & 0.043 \\ 
\hline
\end{tabular}}
\end{table}

\begin{table}
\caption{Sensitivity analysis results: percentage of correct selection with 5 studies for each meta-analysis method.}
\label{tab:results_sens2}
\setlength\tabcolsep{8pt} % default value: 6pt
\scriptsize{\begin{tabular}{l c c c c c c c }
\hline
 & \multicolumn{7}{c}{Dose levels} \\
  & 1 & 2 & 3 & 4 & 5 & 6 & 7   \\
\hline
Scenario 1 & \multicolumn{7}{l}{}  \\
MADF1 & 0.004 & 0.213 & \textbf{0.637} & 0.138 & 0.008 & 0.000 & 0.000 \\ 
  MADF2 & 0.005 & 0.142 & \textbf{0.585} & 0.235 & 0.030 & 0.003 & 0.000 \\ 
  MADF3 & 0.007 & 0.117 & \textbf{0.463} & 0.361 & 0.051 & 0.001 & 0.000 \\ 
  MADF4 & 0.003 & 0.112 & \textbf{0.457} & 0.317 & 0.093 & 0.017 & 0.001 \\ 
 \multicolumn{8}{l}{}  \\  
 Scenario 2 & \multicolumn{7}{l}{}  \\
MADF1 & 0.000 & 0.002 & 0.163 & \textbf{0.796} & 0.038 & 0.001 & 0.000 \\ 
  MADF2 & 0.000 & 0.002 & 0.095 & \textbf{0.817} & 0.084 & 0.002 & 0.000 \\ 
  MADF3 & 0.000 & 0.000 & 0.046 & \textbf{0.801} & 0.152 & 0.001 & 0.000 \\ 
  MADF4 & 0.000 & 0.000 & 0.058 & \textbf{0.766} & 0.170 & 0.005 & 0.001 \\ 
   \multicolumn{8}{l}{}  \\  
 Scenario 3 & \multicolumn{7}{l}{}  \\
MADF1 & 0.000 & 0.000 & 0.011 & 0.333 & \textbf{0.588} & 0.068 & 0.000 \\ 
  MADF2 & 0.000 & 0.000 & 0.003 & 0.206 & \textbf{0.599} & 0.178 & 0.014 \\ 
  MADF3 & 0.000 & 0.000 & 0.001 & 0.132 & \textbf{0.659} & 0.205 & 0.003 \\ 
  MADF4 & 0.000 & 0.000 & 0.001 & 0.116 & \textbf{0.555} & 0.253 & 0.075 \\ 
     \multicolumn{8}{l}{}  \\  
 Scenario 4 & \multicolumn{7}{l}{}  \\
MADF1 & 0.000 & 0.000 & 0.001 & 0.034 & 0.456 & \textbf{0.488} & 0.021 \\ 
  MADF2 & 0.000 & 0.000 & 0.000 & 0.010 & 0.250 & \textbf{0.634} & 0.106 \\ 
  MADF3 & 0.000 & 0.000 & 0.000 & 0.001 & 0.246 & \textbf{0.673} & 0.080 \\ 
  MADF4 & 0.000 & 0.000 & 0.000 & 0.002 & 0.151 & \textbf{0.568} & 0.279 \\
   \multicolumn{8}{l}{}  \\  
 Scenario 5 & \multicolumn{7}{l}{}  \\
MADF1 & 0.000 & 0.019 & 0.284 & \textbf{0.614} & 0.079 & 0.004 & 0.000 \\ 
  MADF2 & 0.000 & 0.015 & 0.187 & \textbf{0.589} & 0.191 & 0.016 & 0.002 \\ 
  MADF3 & 0.000 & 0.011 & 0.097 & \textbf{0.578} & 0.301 & 0.013 & 0.000 \\ 
  MADF4 & 0.000 & 0.009 & 0.104 & \textbf{0.531} & 0.286 & 0.065 & 0.005 \\ 
      \multicolumn{8}{l}{}  \\  
 Scenario 6 & \multicolumn{7}{l}{}  \\
MADF1 & 0.000 & 0.000 & 0.129 & \textbf{0.812} & 0.057 & 0.002 & 0.000 \\ 
  MADF2 & 0.000 & 0.000 & 0.082 & \textbf{0.802} & 0.111 & 0.005 & 0.000 \\ 
  MADF3 & 0.000 & 0.000 & 0.045 & \textbf{0.769} & 0.179 & 0.007 & 0.000 \\ 
  MADF4 & 0.000 & 0.000 & 0.053 & \textbf{0.731} & 0.186 & 0.028 & 0.002 \\
      \multicolumn{8}{l}{}  \\  
 Scenario 7 & \multicolumn{7}{l}{}  \\
MADF1 & 0.000 & 0.000 & 0.013 & 0.317 & \textbf{0.600} & 0.070 & 0.000 \\ 
  MADF2 & 0.000 & 0.000 & 0.004 & 0.170 & \textbf{0.610} & 0.202 & 0.014 \\ 
  MADF3 & 0.000 & 0.000 & 0.001 & 0.123 & \textbf{0.658} & 0.214 & 0.004 \\ 
  MADF4 & 0.000 & 0.000 & 0.002 & 0.104 & \textbf{0.516} & 0.290 & 0.088 \\
     \multicolumn{8}{l}{}  \\  
 Scenario 8 & \multicolumn{7}{l}{}  \\
MADF1 & 0.000 & 0.000 & 0.042 & 0.413 & \textbf{0.500} & 0.044 & 0.001 \\ 
  MADF2 & 0.000 & 0.000 & 0.015 & 0.232 & \textbf{0.554} & 0.181 & 0.018 \\ 
  MADF3 & 0.000 & 0.001 & 0.005 & 0.150 & \textbf{0.645} & 0.192 & 0.007 \\ 
  MADF4 & 0.000 & 0.000 & 0.005 & 0.141 & \textbf{0.514} & 0.248 & 0.092 \\ 
     \multicolumn{8}{l}{}  \\  
 Scenario 9 & \multicolumn{7}{l}{}  \\
MADF1 & 0.000 & 0.000 & 0.008 & 0.271 & \textbf{0.635} & 0.085 & 0.001 \\ 
  MADF2 & 0.000 & 0.000 & 0.005 & 0.173 & \textbf{0.622} & 0.189 & 0.011 \\ 
  MADF3 & 0.000 & 0.000 & 0.002 & 0.114 & \textbf{0.684} & 0.196 & 0.004 \\ 
  MADF4 & 0.000 & 0.000 & 0.005 & 0.095 & \textbf{0.552} & 0.279 & 0.069 \\ 
\hline
\end{tabular}}
\end{table}

  \clearpage
  \bibliographystyle{abbrvnat}
  \bibliography{biblio_dosemeta}

\end{document}